\documentclass[fleqn,twoside]{article}
\usepackage{espcrc2}
\usepackage{graphicx}
\usepackage[figuresright]{rotating}
\newcommand{\quq}{\theta_{13}}
\newcommand{\dmsq}{\Delta m^2}
\title{New Reactor Neutrino Experiments        besides Double-CHOOZ}
\author{M. Goodman\address[HEP]{High Energy Physics Division
\\HEP 362 \\ Argonne Illinois 60439}}
\begin{document}
\begin{abstract}
Several new reactor neutrino experiments are being considered to
measure the parameter $\quq$.  The current plans for Angra, Braidwood,
Daya Bay, KASKA and KR2DET are reviewed.  A case is made that, together
with Double-CHOOZ, a future world program should include at least three
such experiments.
\end{abstract}
\maketitle
\section{Introduction and Remarks about $\quq$}
\par The mixing angle $\quq$ is the last mixing angle to be measured.
A non-zero value of $\quq$ is key to the possibility of measuring 
CP violation in future long-baseline accelerator experiments.  The best
limit on $\quq$ comes from the reactor experiment CHOOZ.\cite{bib:chooz}
It has recently been recognized that the best possibility for rapidly 
measuring a non-zero value of the parameter $\quq$ comes from
a new reactor experiment using two or more identical detectors and
improved control of systematic errors.
\par
During 2003, a group of neutrino physicists assembled for a series of 
workshops to explore the capabilities of a new nuclear reactor experiment.
Together, they wrote a white paper called
``A New Nuclear Reactor $\nu$ Experiment to Measure $\quq$"\cite{bib:white}.
Several issues were addressed in this document:
\begin{itemize}
\item [$\odot$] The optimal baseline distances, luminosity scaling and the impact
of systematics
\item [$\odot$] Previous reactor Experiments
\item [$\odot$] Detector Design
\item [$\odot$] Calibration requirements and procedures
\item [$\odot$] Detector overburden and backgrounds
\item [$\odot$] Systematic error budget
\item [$\odot$] Possible sites
\item [$\odot$] Other physics that could be addressed in a new experiment
\item [$\odot$] Tunneling issues
\item [$\odot$] Safety
\item [$\odot$] Outreach and Education
\end{itemize}
The document was written with a site-independent description of such a new
experiment, but seven appendices were included about particular initiatives.
  In a companion presentation at this meeting, the 
Double-CHOOZ project was described.\cite{bib:mention}  That project may be 
the first one
underway, but due to limitations of its size and distance, more sensitive
and larger but later projects are also under active development.  This paper
will describe the plans for some of these ``other" experiments.
\par
\begin{figure}[htb]
\includegraphics[width=3.0in]{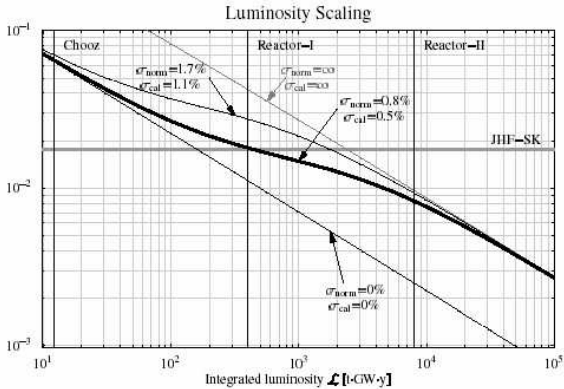}
\caption{ 
A plot of the $\sin^2 (2\quq)$ sensitivity (vertical) versus reactor 
luminosity (horizontal) for particular detector
distances, systematic error assumptions 
and $\dmsq$ from Reference \cite{bib:huber}.}
\label{fig:huber}
\end{figure}

 The possibility of a new reactor experiment to measure $\quq$ was a key
concern of the DNP/DPB/DAP/DPB Joint Study on the Future of Neutrino Physics,
which recently issued the report, ``The Neutrino Matrix"\cite{bib:aps}.  One of the two
high priority recommendations was for ``a comprehensive U.S. program to complete
our understanding of neutrino mixing, to determine the character of the neutrino
mass spectrum and to search for CP violation among neutrinos," including
``an expeditiously deployed multi-detector reactor experiment with sensitivity
to $\bar{\nu}_e$ disappearance down to $\sin^2 2\quq = 0.01$, an order of
magnitude below present limits."  The choice of 0.01 reflects the desire to
provide unambiguous information about $\quq$ that would be useful to
foreseeable future long-baseline experiments that could measure CP violation
and matter effects.  As shown in Figure \ref{fig:huber}, it also reflects
the approximate foreseeable sensitivity for the next generation or two of
reactor experiments.

\section{Sites}
\subsection{KR2DET}
The first idea for a new nuclear reactor experiment was the KR2DET project
using a reactor at Krasnoyarsk in Russia which 
utilizes a rapid fuel cycle.\cite{bib:kr2det}
  Use of
a rapid fuel cycle would reduce the systematic error on the flux of neutrinos
due to burn up.  The most attractive feature of this experiment is that the
entire site is located at a depth of 600 meters of water equivalent (mwe).
They proposed two detectors, each a 4.7 m diameter liquid scintillator target,
enclosed in a transparent spherical balloon.  Each detector would use 
800 8-inch photomultipliers, mounted on a stainless steel screen 
which separates a veto from the inner detector.  
\par The existence of the underground lab together with the reactor is a strong
advantage. But this particular reactor will
be closed by treaty before the year 2012, because it is of the type that creates weapons
grade material; indeed that is what it was built for.  It is currently used
to provide power for the city of Krasnoyarsk, and will be replaced with an
energy plant which cannot be used to create Plutonium.  It might have
been possible to complete a 300 GW-ton-year experiment before the reactor
turned off.  However a planned visit by western physicists in 2003 was 
canceled by local security officials, and development of this experiment
effectively ended at that time.
\subsection{Diablo Canyon}
The idea for using the Diablo Canyon reactor has been developed by the Berkeley 
group and others.\cite{bib:diablo}  They did a geological evaluation
and tunnel cost estimate for two detectors near this pair of cores with
a total 6.1 GW-thermal power.  The excavation of a horizontal tunnel in the
coastal mountains can provide overburden up to 800 mwe with tunnel distances
up to 3 km.  
\par Originally the power company, PG\&E, cooperated in the development of the
proposal.  Construction of an on-site waste disposal facility
provided a window of time
during which there would be civil construction on site anyway, and the
addition of a tunnel would not create additional environmental concerns.
However, for reasons that are not clear, 
PG\&E decided that this project would
not be in its interest, and they have
discouraged further development of this
proposal.  
\subsection{KASKA}
The largest concentration of nuclear power plants in the world, and
hence the most powerful neutrino source, is at
Kashiwazaki-Kariwa, with a total of 24.3 GW-thermal.  Seven
reactors are located
along the west coast of Japan.  The reactors are located
in one cluster of four and one cluster of three.  A favorable geometry
exists which consists of two near detectors and one far detector, located
on the site boundary.  The ground is soft, so rather than constructing
a cavern underground, a detector would be placed in a shaft-hole,
drilled with the techniques used for making bridges.
The experiment, called KASKA, plans to use four volumes of liquid:
\begin{enumerate}
\item An inner Neutrino target consisting of Gadolinium loaded liquid
scintillator.
\item A gamma-catcher consisting of scintillator without Gd.
\item A buffer region consisting of mineral oil without scintillator.
\item An optical barrier before a cosmic ray veto region with weak
scintillator.
\end{enumerate}
KASKA recently received some R\&D funding to drill a test hole and
make some preliminary cosmic ray studies.  They are also developing
a prototype detector.
\subsection{Braidwood}
The Braidwood project is an active collaboration 
of Chicago, Oxford, Kansas State, Columbia, ANL, FNAL, BNL, Pittsburgh,
MIT, Michigan and Texas.
The Braidwood
reactor is a two unit facility located about 50 miles southwest of Chicago.
An R\&D proposal for one year has been submitted to the NSF and the DOE and
a full proposal is expected in about a year.  The baseline design consists
of two near and two far detectors with a 3.5 m outer radius.  The distances
will be along the center line of the two reactor cores, and approximately
200 m and 1500 m from the cores.  The planned depth is 180 m corresponding to
450 mwe.  The design currently is for a two zone detector, using 
0.1\% Gd loaded scintillator in the inner volume and mineral oil in the outer
volume.  The radius of the acrylic sphere will be 2.6 m.
There will be approximately 25\% phototube coverage, using 1000 PMTs per
detector.  The detectors will be movable.  Baseline sensitivities will not
use the reduced systematic from moving, but this will
be used as a cross-check.
A veto system of approximately 1 m of passive shielding surrounded by active
veto counters on the top and sides will be used.  There will also be a layer
of active veto counters below the detectors.

\subsection{Daya Bay}
The Daya Bay complex of reactors is located in China outside Hong Kong.
There are two reactors at Daya Bay, two more at Ling Ao, and plans for
another two nearby at Ling Dong.  Nearby are large hills.  Two near
detectors would be built, one near Daya Bay and the other between
Ling Ao and Ling Dong.  A far detector could be located deep in the
hills under a considerable overburden of 1200 mwe.  When Ling Dong is
complete, this will be the second most powerful reactor complex in the
world.  
\subsection{Angra dos Rois}
Angra dos Reis is located about 150 km south of Rio de Janeiro in 
Brazil.  The nuclear
facility contains two operational reactors.  The Angra-I reactor is an older low power
reactor (1.5 GW-th).  Angra-II is 4.1 GW-th.  The reactors are located
on the coast and the reactor company controls a strip of land that
stretches inland about 1-1.5 km and is approximately 4 km along the coast.  
Much of the terrain is mountainous granite with multiple peaks up to 600 m
high.  The focus of
the design work is now oriented toward large detectors, larger than 
100 ton each, in order to take advantage of the spectral
shape test.  This means
that energy calibration will be crucial.  The next workshop of the International
Working Group is tentatively scheduled for 23-25 February 2005 near Angra,
Brazil.
\subsection{Common concerns}
The experiences at Diablo Canyon and Krasnoyarsk
emphasizes that 
there has been a history of less than stellar cooperation between 
reactor power companies/authorities and reactor neutrino experiments.
The cases are all different, but include  San Onofre,  Palo Verde, 
 Krasnoyarsk, and  Diablo Canyon.   An appreciation of the needs and 
concerns of the reactor authorities is crucial in developing a new 
larger program.   These issues include security, safety and economics.
So far, the other reactor operators:  Electricite de France,
Tokyo Electric Power, Exelon,
and the governments of China and Brazil have been more willing to
cooperate with the physicists investigating new experiments.
Careful and continued consideration to these concerns is crucial for
a successful ambitious future program of neutrino reactor experiments.

\begin{table*}[hbtp] 
\centering 
{\begin{tabular}{|l|l|c|c|c|} 
\hline 
Proposal & Power & Baseline & Detector & Overburden \\ \hline
         & (GW)  & Near/Far (m)& Near/Far (t) & Near/far (mwe) \\ \hline
Angra dos Reis (Brazil) & 4.1 & 300/1300 & 25/25 & 60/600 \\ \hline
Braidwood (US) & 6.5 & 200/1500 & 25/50 & 250/250 \\ \hline
Double-CHOOZ (France) & 8.4 & 150/1050 & 10/10 & 50/300 \\ \hline
Daya Bay (China) & 11.6 & 300/1500 & 25/50 & 200/1000 \\ \hline
Diablo Canyon (US) & 6.4 & 400/1800 & 25/50 & 100/700 \\ \hline
KASKA (Japan) & 24.3 & 300/1300 & 8.5/8.5 & 140/600 \\ \hline
KR2DET (Russia) & 3.2 & 115/1000 & 46/46 & 600/600 \\ \hline
\hline 
\end{tabular}} 
\caption{Detector parameters for proposed experiments.  Note that except
for the reactor powers, and the Double-Chooz far detector distance and
overburden, and the KR2DET overburden, all parameters are subject to
design and optimization.}\label{tab:comp} 
\end{table*}
\subsection{Comments}
A summary of proposed experiment parameters is given in Table \ref{tab:comp}.
As funding agencies try to evaluate the best strategy, it is important to
keep in mind that except for power, most entries there are subject to
design and optimization.  
The optimum strategy for a future program of reactor neutrino experiments
includes consideration of detector sizes and systematic errors, but also
costs and institutional and national considerations, not fully
considered in this
presentation.  
\par A glance at Figure \ref{fig:huber} indicates some of the
challenges.  The thick curve represents the sensitivity/luminosity relation
using a reasonable guess for 
the ultimate systematic errors of a new experiment.\cite{bib:huber}
Note that the luminosity (in GW-t-years) for a sensitivity of 0.01 is 70 times larger than the
luminosity for a sensitivity of 0.03, which is the goal for Double-CHOOZ.
I think a first experiment such as Double-CHOOZ is a prerequisite for
the more sensitive experiment, whether it meets its goal by increased luminosity
or by decreased systematic errors.  But I also think the story won't stop
there.   Just as experiments to increase knowledge of the elements of the
CKM matrix are continuing, better knowledge about $\quq$ will be a continued
goal of the particle physics community for some time to come, whether or not
a non-zero value is found soon.  I foresee at least three new rounds of 
neutrino reactor experiments, each building on the lessons of the
previous experiments.

\end{document}